# Energy transfer rate in double-layer graphene systems: Linear regime

B. Bahrami

*Department of Physics, Shahid Beheshti University, G. C., Evin, 1983963113, Tehran, Iran;*

*b_bahrami@sbu.ac.ir*

T. Vazifehshenas*

*Department of Physics, Shahid Beheshti University, G. C., Evin, 1983963113, Tehran, Iran;*

*t-vazifeh@cc.sbu.ac.ir*

* Corresponding author: Tel:+98 21 29902784; Fax: +98 21 22431666; E-mail: t-vazifeh@cc.sbu.ac.ir

**Abstract**

We investigate theoretically the energy transfer phenomenon in a double-layer graphene (DLG) system in which two layers are coupled due to the Coulomb interlayer interaction without appreciable interlayer tunneling. We use the balance equation approach and the dynamic and temperature dependent random phase approximation (RPA) screening function in our calculations to obtain the rates of energy transfer between two graphene layers at different layer electron temperatures, densities and interlayer spacings and compare the results with those calculated for the conventional double-layer two-dimensional electron gas (2DEG) systems. In addition, we study the effect of changing substrate dielectric constant on the rate of energy transfer. The general behavior of the energy transfer rate in the DLG is qualitatively similar to that obtained in the double-layer 2DEG but quantitatively its DLG values are an order of magnitude greater. Also, at large electron temperature differences between two layers, the electron density dependence of the energy transfer for the DLG system is significantly different from that found for the double-layer 2DEG system, particularly in case of unequal layer electron densities.

# 1. Introduction

Graphene is a two-dimensional honeycomb structure of carbon atoms composed of two triangular sublattices. The single-layer graphene (SLG) is a gapless semiconductor and has unusual massless and chiral carriers near the Dirac points where its energy spectrum is linear. The unique features of graphene have been attracted a great deal of interest theoretically and experimentally and made it a good candidate for technological applications [1-7].

The double-layer graphene (DLG) system is formed by placing two parallel SLG close together to be coupled through the Coulomb interaction but far enough to prevent the electron tunneling. In the DLG structures as well as the conventional double-layer two-dimensional electron gas (2DEG) systems, the interlayer interaction plays an important role [8-12] and results in some novel phenomena such as the momentum transfer (Coulomb drag) and energy transfer between two layers without exchanging any carrier. The Coulomb drag in a double-layer system in which a current driven through one layer induces a voltage difference in the other layer provides a unique approach to investigate the electron-electron interaction directly in a low-dimensional many-body system from a transport measurement [13-15]. The various aspects of this effect have been studied in detail, in several experimental and theoretical works on double-layer 2DEG [16-25] and DLG [26-31] systems.

By allowing the electron temperature difference between two adjacent layers in double-layer systems, the energy of electrons in one layer can be transferred to the electrons in another, located in close well-separated proximity while the tunneling effect is prohibited. This hot electron transport process can be occurred with or without an externally applied electric field [16, 31] called non-linear and linear regimes, respectively. The balance equation method which is a powerful tool in describing the steady state response of nanostructures to the external fields has successfully applied to a variety of situations involving transport phenomena [32-45]. Based on the energy balance formalism, the rate of energy transfer in the double quantum systems has been investigated in a few papers and its dependence on the electron temperature, electron density and separation between two components has been reported [46-49]. Specially, Tanatar considered Coulomb coupled quantum wire [50] and well [51] systems and calculated the effects of the static and dynamic screenings within the random phase approximation (RPA) on the energy transfer rate both in linear and non-linear regimes.

In this work we investigate the energy transfer rate in a DLG system employing the general (non-parabolic) energy band balance equation theory suggested by Lei et al [36-38] in linear regime. We use the temperature dependent dynamic RPA dielectric function for the screened interlayer interaction in our calculations. We obtain the energy transfer rate between two layers with the same electron density as functions of temperature for the different layer densities and interlayer spacings. For comparison, we also show the corresponding results for the double-layer 2DEG systems. In addition, the effect of substrate and layer density imbalance on the power transferred between two graphene layers are taken into account.

The article is organized as follows. In next section we describe the model and summarize the theory of energy transfer rate for a DLG system in framework of the balance equation approach. We also present the screened function of a DLG structure within the RPA formalism in this section. The section 3 is dedicated to numerical results and discussion. Finally, in section 4 a conclusion of our work is given.

## 2. Theory

We consider two parallel n-type doped graphene layers which are separated by a distance $d$ and have the electron densities, $n_1$ and $n_2$ and the electron temperatures $T_1$ and $T_2$. In each layer the electron density is related to the Fermi wave vector, $k_F$, as $n = k_F^2/\pi$ and to the Fermi temperature, $T_F$, by $n = (T_F/v_F)^2/\pi$ where $v_F = 10^6$ m/s is the Fermi velocity. Throughout this paper, we put $\hbar = K_B = 1$. We also assume this two graphene layers are placed close together to have an effective interlayer interaction but far enough that there is no electron tunneling between them.

In framework of the balance equation approximation for arbitrary energy bands, the transport properties of electrons in the presence of an uniform electric field are obtained in terms of the average lattice momentum shift, $p_a$, electron temperature, $T_a$, and the chemical potential, $\mu_a$, of each layer $(a=1,2)$. In the limit of weak electric field where $p_a = 0$, the rate of energy transfer between two layers in a DLG system is obtained from [41]:

$$P_{12} = g_s g_v \sum_{\substack{\mathbf{k},\mathbf{k_1},\mathbf{q} \\ \lambda,\lambda',\lambda_1,\lambda_1'}} |v_{12}(\mathbf{k},\mathbf{k_1},\mathbf{q})|^2 [E(\mathbf{k},\lambda) - E(\mathbf{k}+\mathbf{q},\lambda')] \Upsilon_{12}(\mathbf{k}\lambda\lambda',\mathbf{k_1}\lambda_1\lambda_1',\mathbf{q}) \quad (1)$$

where $g_s = 2$ and $g_v = 2$ being the spin and valley degeneracies, $\mathbf{k}$, $\mathbf{k_1}$ and $\mathbf{q}$ are the two-dimensional wave vectors in plane of graphene layer and $\Upsilon_{12}$ is given by:

$$\Upsilon_{12}(\mathbf{k}\lambda\lambda',\mathbf{k_1}\lambda_1\lambda_1',\mathbf{q}) = \pi g_s g_v \frac{\delta\left[E(\mathbf{k},\lambda) - E(\mathbf{k}+\mathbf{q},\lambda') + E(\mathbf{k_1}+\mathbf{q},\lambda_1') - E(\mathbf{k_1},\lambda_1)\right]}{\left|\varepsilon^{RPA}(\mathbf{q},E(\mathbf{k},\lambda) - E(\mathbf{k}+\mathbf{q},\lambda'),T_1,T_2)\right|^2}$$
$$\times\left[f(\frac{\xi_{1\mathbf{k},\lambda}}{T_1}) - f(\frac{\xi_{1\mathbf{k+q},\lambda'}}{T_1})\right]\left[f(\frac{\xi_{2\mathbf{k_1},\lambda_1}}{T_2}) - f(\frac{\xi_{2\mathbf{k_1+q},\lambda_1'}}{T_2})\right]\left[n(\frac{E(\mathbf{k},\lambda) - E(\mathbf{k}+\mathbf{q},\lambda')}{T_1}) - n(\frac{E(\mathbf{k_1},\lambda_1) - E(\mathbf{k_1}+\mathbf{q},\lambda_1')}{T_2})\right]$$
(2)

Here $\varepsilon^{RPA}$ is the RPA screening function, $n(x) = 1/[\exp(x) - 1]$ and $f(x) = 1/[\exp(x) + 1]$ are the Bose and Fermi-Dirac distribution functions, respectively. In the above equations, we have defined $\xi_{a\mathbf{k},\lambda} = E(\mathbf{k},\lambda) - \mu_a$ and $E(\mathbf{k},\lambda) = \lambda v_F k$ with $\lambda = 1$ and $\lambda = -1$ refer to the valance and conduction bands and $\mu_a$ being the chemical potential of $a$ th layer which is obtained from the normalization condition:

$$n_a = \int_{-\infty}^{\infty} dE \frac{D(E)}{1 + e^{(E-\mu_a)/T_a}} \tag{3}$$

with $D(E) = 2E/\pi v_F^2$. The interlayer interaction, $v_{12}(\mathbf{k},\mathbf{k_1},\mathbf{q})$, is given by:

$$v_{12}(\mathbf{k},\mathbf{k_1},\mathbf{q}) = V_{12}(\mathbf{q}) g_{11}(\mathbf{k},\mathbf{q}) g_{22}(\mathbf{k_1},\mathbf{q}) \tag{4}$$

where $V_{ab}(\mathbf{q}) = (2\pi e^2/\kappa q) \exp[-qd(1-\delta_{ab})]$ is the Fourier transform of the bare Coulomb potential with $\kappa$ arising from the effective background dielectric constant, $\kappa = (1+\varepsilon)/2$, and $\varepsilon$ is the substrate dielectric constant. The graphene form factor $g_{11}(\mathbf{k},\mathbf{q}) = g_{22}(\mathbf{k},\mathbf{q}) = g(\mathbf{k},\mathbf{q},\lambda,\lambda')$ is defined as [45]:

$$g(\mathbf{k},\mathbf{q},\lambda,\lambda') = \langle \mathbf{k}+\mathbf{q},\lambda' | \mathbf{k},\lambda \rangle = \frac{1}{2}\left[1 + \lambda\lambda' \exp(\phi_\mathbf{k} - \phi_\mathbf{k+q})\right] \tag{5}$$

with $\phi_\mathbf{k} = Arc\tan(k_y/k_x)$.

After some algebra, Eq. (1) can be rewritten as (see the Appendix):

$$P_{12} = -\sum_\mathbf{q} \int_{-\infty}^{+\infty} \frac{d\omega}{\pi} \omega \left|\frac{V_{12}(\mathbf{q})}{\varepsilon^{RPA}(\mathbf{q},\omega,T_1,T_2)}\right|^2 \left[n(\frac{\omega}{T_1}) - n(\frac{\omega}{T_2})\right] \text{Im}\chi_1(\mathbf{q},\omega,T_1) \text{Im}\chi_2(-\mathbf{q},-\omega,T_2) \tag{6}$$

where $\text{Im}\chi_a(\mathbf{q},\omega,T_a)$ is the imaginary part of temperature dependent non-interacting polarizability of the $a$ th graphene layer [12]:

$$\chi_a(\mathbf{q},\omega,T_a) = g_s g_v \lim_{\eta \to 0} \sum_{\lambda,\lambda'=\pm 1} \int \frac{d^2\mathbf{k}}{(2\pi)^2} \left[\left(\frac{1 + \lambda\lambda' \cos\theta}{2}\right) \frac{\left[f(\xi_{a\mathbf{k+q},\lambda'}/T_a) - f(\xi_{a\mathbf{k},\lambda}/T_a)\right]}{\omega + E(\mathbf{k}+\mathbf{q},\lambda') - E(\mathbf{k},\lambda) + i\eta}\right] \tag{7}$$

where $\theta$ is the angle between $\mathbf{k}$ and $\mathbf{k}+\mathbf{q}$. The RPA screening function at finite temperature for a double-layer system, $\varepsilon^{RPA}(\mathbf{q},\omega,T_1,T_2)$, is given by [10]:

$$\varepsilon^{RPA}(\mathbf{q},\omega,T_1,T_2) = [1 - V_{11}(\mathbf{q})\chi_1(\mathbf{q},\omega,T_1)][1 - V_{22}(\mathbf{q})\chi_2(\mathbf{q},\omega,T_2)] - V_{12}^2(\mathbf{q})\chi_1(\mathbf{q},\omega,T_1)\chi_2(\mathbf{q},\omega,T_2) \qquad (8)$$

Using the above equations, we can calculate the energy transfer rate in a DLG system.

## 3. Results and discussion

We have performed numerical calculations based on the balance equation theory to obtain the energy transfer rate, $P_{12}$, for a DLG system in the linear regime. We have accounted for the screening effects by using the temperature dependent dynamic RPA dielectric function for a two-component system. We have studied the dependence of the energy transfer rate on the layer density, electron temperature, separation between two layers and layer substrate in a DLG system with two identical layers. Also, the density imbalance effect on the power transferred from the hot layer to the cold layer has been investigated. In addition, we have compared our result with those calculated for a double-layer 2DEG system with zero thickness layers, $w=0$. In the following figures for the DLG system, we take $\kappa = 2.45$ unless otherwise stated.

The energy transfer rates between two identical layers ($n_1 = n_2 = n$) as functions of dimensionless electron temperature in layer 2, $T_2/T_F$, for three distinct values of the electron densities $n = 2$, $1$ and $0.5 \times 10^{12} \text{cm}^{-2}$ in the DLG and double-layer 2DEG systems are shown in Figs. 1(a) and 1(b), respectively. The two layers are separated a distance $d = 20\text{nm}$ and the electron temperature in the layer 1 is kept at $T_1 = T_F$. As it is observed, the energy transfer rate in a DLG system presents a similar behavior as in the double-layer 2DEG system but its DLG values are an order of magnitude greater. This can be explained by the fact that, because of the zero energy band-gap and linear energy dispersion relation of the graphene, the screened interlayer potential in a DLG is stronger than a double-layer 2DEG so that more power can be transferred between two graphene layers.

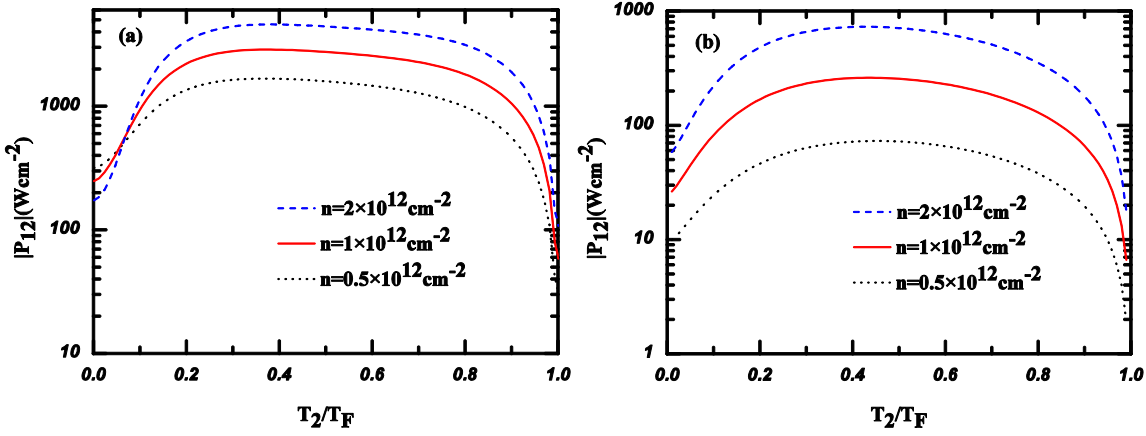

**Fig. 1.** The energy transfer rate as functions of dimensionless electron temperature in layer 2, $T_2/T_F$, for three different electron densities in the (a) DLG and (b) double-layer 2DEG systems. Both systems consist of two identical layers which are separated a distance $d = 20\text{nm}$. The temperature in layer 1 is kept constant at $T_1 = T_F$.

According to Fig. 1(a) at low electron temperatures in layer 2, $T_2/T_F < 0.1$, the energy transfer rate in the DLG has different dependence on the electron density and faster growth with respect to the double-layer 2DEG. A it is expected, the energy transfer rate vanishes in all cases when the temperatures of both layers are equal.

The effect of interlayer spacing on $P_{12}$ in both double-layer systems with identical layer electron densities, $n = 1 \times 10^{12}\text{cm}^{-2}$, are depicted in Figs. 2(a) and 2(b). Here, the results are obtained for three different interlayer distances, $d = 20, 30$ and $50\text{nm}$, and the electron temperature of the layer $T_1 = T_F$. In spite of an order of magnitude increase in energy transfer rate values, again the DLG system shows the same behavior as the double-layer 2DEG system; by increasing the distance between two layers, the rate of energy transfer from the higher temperature layer to the lower temperature one decreases. This simply is a consequence of weakening the interlayer interaction arising from more electrical isolation of layers.

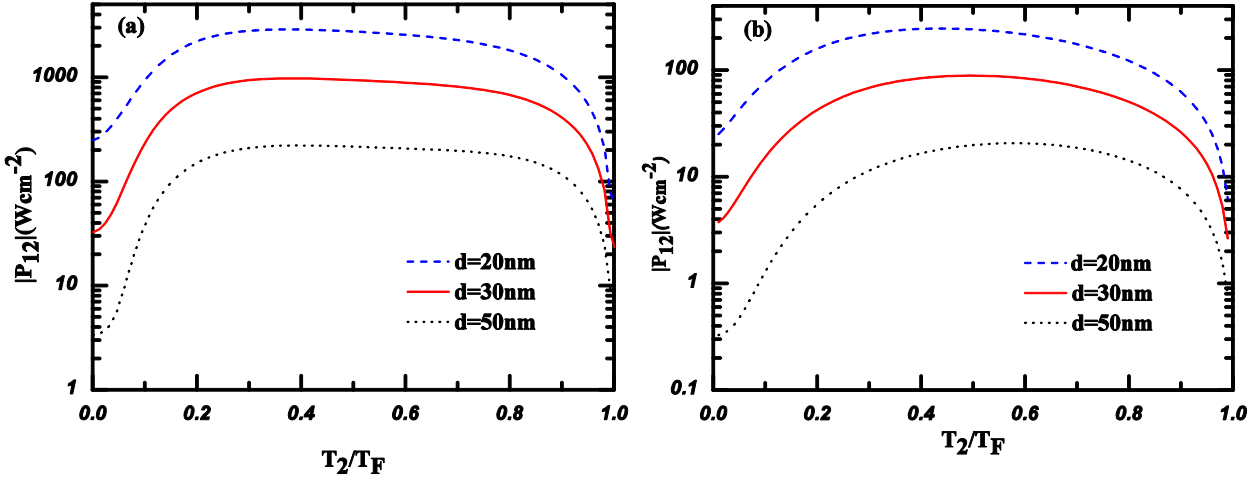

**Fig. 2.** The energy transfer rate as functions of dimensionless electron temperature in layer 2, $T_2/T_F$, for three different interlayer spacings in the (a) DLG and (b) double-layer 2DEG systems. Both systems consist of two identical layers with the same layer density, $n = 1\times10^{12}\,\text{cm}^{-2}$. The temperature in layer 1 is kept constant at $T_1 = T_F$.

Fig. 3 demonstrates the energy transfer rate in the DLG structures with two different substrates as a function of $T_2/T_F$ when the two layers have equal density, $n = 1\times10^{12}\,\text{cm}^{-2}$, and for the case $T_1 = T_F$ and $d = 20\,\text{nm}$. In this figure, the influence of the substrate is taken into account by introducing the dimensionless parameter $r_s = e^2/\kappa \upsilon_F$, the ratio of the bare Coulomb potential strength to the single-particle kinetic energy.

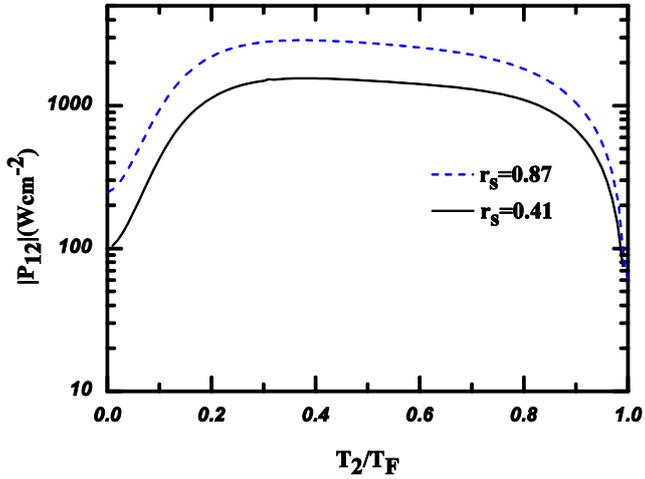

**Fig. 3.** The energy transfer rate as functions of dimensionless electron temperature in layer 2, $T_2/T_F$, for the DLG system with two different substrates. The system consists of two identical layers with the same layer density, $n = 1\times10^{12}\,\text{cm}^{-2}$. The temperature in layer 1 is kept constant at $T_1 = T_F$. The upper and lower curves correspond to the $SiO_2$ and $SiC$ substrates, respectively.

Here, the energy transfer rate is plotted for two distinct values of $r_s = 0.87$ and $0.41$ corresponding to the $\varepsilon_{SiO_2} = 3.9$ and $\varepsilon_{SiC} = 9.8$, respectively [53]. Since $r_s$ and the substrate dielectric constant are inversely related, the energy transfer rate decreases by increasing the background dielectric constant of graphene.

The effect of different electron density of two layers on the energy transfer rate both in the DLG and double-layer 2DEG systems is illustrated in Figs. 4(a) and 4(b). We assume $n_1 = 1 \times 10^{12} \text{cm}^{-2}$, $T_1 = T_{F1}$ and allow $n_2 / n_1$ to vary from $0.2$ to $2$ for five different $T_2$ values.

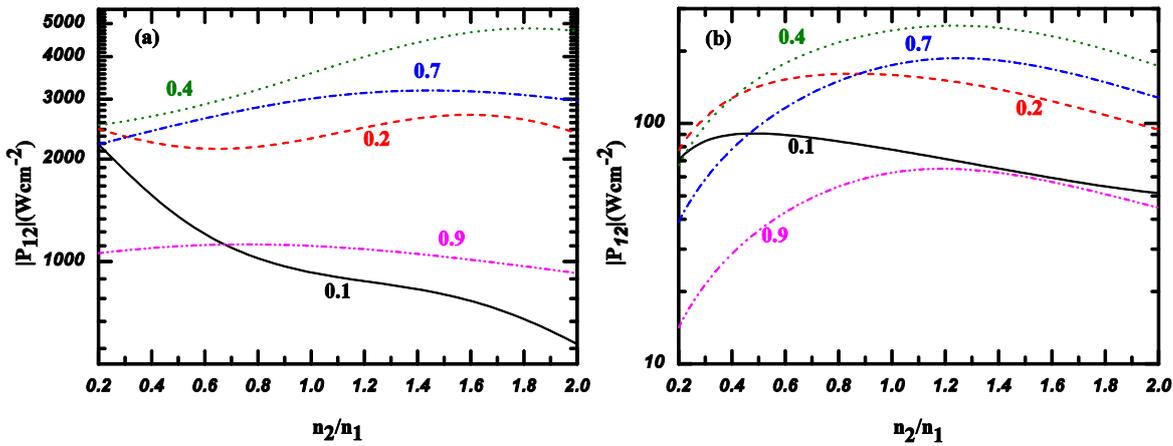

**Fig. 4.** The energy transfer rate as functions of the ratio layer electron densities $n_2 / n_1$ at five different electron temperatures in layer 2: $T_2/T_{F1} = 0.1, 0.2, 0.4, 0.7$ and $0.9$ for the (a) DLG and (b) double-layer 2DEG systems with an interlayer separation $d = 20$nm. The electron density and temperature of layer 1 are kept constants at $n_1 = 1 \times 10^{12} \text{cm}^{-2}$ and $T_1 = T_{F1}$, respectively.

As it can be seen from Fig. 4(a), the energy transfer rate in the DLG is much more sensitive to the layer electron densities ratio than in the double-layer 2DEG and shows there is no general behavior specially at low electron temperature in layer 2. On the other hand, at high $T_2 / T_F$ or small temperature difference between two layers, the behavior of the transferred power in the DLG is comparable to the conventional double-layer 2DEG and both exhibit a peak. This peak shifts to smaller $n_2 / n_1$ values with increasing the electron temperature in layer 2.

## 4. Conclusion

In conclusion, we have studied the energy transfer phenomenon in a DLG system using the balance equation approach in linear regime. We employ the full dynamic and temperature dependent RPA screening function in our calculations to obtain the energy transfer rate in linear regime from the hot layer to the cold one at different electron densities and layer spacings and compare with the results of previous work on the conventional double-layer 2DEG systems. In general, the energy transfer rate in a DLG system shows the same behavior as the double-layer 2DEG but exhibits an order of magnitude larger values. We also study the change of $P_{12}$ by altering the substrate dielectric constant. Furthermore, the case of layers with unequal electron densities is taken into account and it is found that the energy transfer rate for DLG system is more sensitive to the density imbalance than the double-layer 2DEG system.

**Appendix: Calculation of $P_{12}$**

In this appendix we give the details of the calculation yielding Eq. (6). Starting from Eq. (1) and using the following relation in Eq. (2):

$$\delta[E(\mathbf{k},\lambda) - E(\mathbf{k}+\mathbf{q},\lambda') + E(\mathbf{k_1}+\mathbf{q},\lambda_1') - E(\mathbf{k_1},\lambda_1)]$$
$$= \int_{-\infty}^{+\infty} d\omega \, \delta[\omega + E(\mathbf{k}+\mathbf{q},\lambda') - E(\mathbf{k},\lambda)] \delta[\omega + E(\mathbf{k_1}+\mathbf{q},\lambda_1') - E(\mathbf{k_1},\lambda_1)] \quad (A.1)$$

we can write

$$P_{12} = \pi (g_s g_v)^2 \sum_{\substack{\mathbf{k},\mathbf{k_1},\mathbf{q} \\ \lambda,\lambda',\lambda_1,\lambda_1'}} |V_{12}(\mathbf{q})|^2 |g_{11}(\mathbf{k},\mathbf{q})|^2 |g_{22}(\mathbf{k_1},\mathbf{q})|^2 [E(\mathbf{k},\lambda) - E(\mathbf{k}+\mathbf{q},\lambda')]$$
$$\times \int_{-\infty}^{+\infty} d\omega \frac{\delta[\omega + E(\mathbf{k}+\mathbf{q},\lambda') - E(\mathbf{k},\lambda)] \delta[\omega + E(\mathbf{k_1}+\mathbf{q},\lambda_1') - E(\mathbf{k_1},\lambda_1)]}{|\varepsilon^{RPA}(\mathbf{q}, E(\mathbf{k},\lambda) - E(\mathbf{k}+\mathbf{q},\lambda'), T_1, T_2)|^2} \quad (A.2)$$
$$\times \left[ f(\frac{\xi_{1\mathbf{k},\lambda}}{T_1}) - f(\frac{\xi_{1\mathbf{k}+\mathbf{q},\lambda'}}{T_1}) \right] \left[ f(\frac{\xi_{2\mathbf{k_1},\lambda_1}}{T_2}) - f(\frac{\xi_{2\mathbf{k_1}+\mathbf{q},\lambda_1'}}{T_2}) \right] \left[ n(\frac{E(\mathbf{k},\lambda) - E(\mathbf{k}+\mathbf{q},\lambda')}{T_1}) - n(\frac{E(\mathbf{k_1},\lambda_1) - E(\mathbf{k_1}+\mathbf{q},\lambda_1')}{T_2}) \right]$$

By using the graphene form factor relation

$$|g_{11}(\mathbf{k},\mathbf{q})|^2 = \frac{1}{4}(1 + \lambda\lambda' e^{i(\phi_\mathbf{k} - \phi_{\mathbf{k}+\mathbf{q}})})(1 + \lambda\lambda' e^{-i(\phi_\mathbf{k} - \phi_{\mathbf{k}+\mathbf{q}})}) = \frac{1 + \lambda\lambda' \cos\theta}{2} \quad (A.3)$$

and the imaginary part of the temperature dependent non-interacting polarizability for a graphene layer:

$$\mathrm{Im}\,\chi_a(\mathbf{q},\omega,T) = g_s g_v \pi \sum_{\lambda,\lambda'=\pm 1} \int \frac{d^2\mathbf{k}}{(2\pi)^2} \left[ \left(\frac{1 + \lambda\lambda'\cos\theta}{2}\right) \left[ f(\frac{\xi_{a\mathbf{k}+\mathbf{q},\lambda'}}{T_a}) - f(\frac{\xi_{a\mathbf{k},\lambda}}{T_a}) \right] \delta[\omega + E(\mathbf{k}+\mathbf{q},\lambda') - E(\mathbf{k},\lambda)] \right] \quad (A.4)$$

and substituting them into the Eq. (A2), we finally obtain the following expression for the energy transfer rate in a DLG system:

$$P_{12} = \sum_{\mathbf{q}} \int_{-\infty}^{+\infty} \frac{d\omega}{\pi} \omega \left| \frac{V_{12}(\mathbf{q})}{\varepsilon^{RPA}(\mathbf{q},\omega,T_1,T_2)} \right|^2 \left[ n(\frac{\omega}{T_1}) - n(\frac{\omega}{T_2}) \right] \operatorname{Im}\chi_1(\mathbf{q},\omega,T_1) \operatorname{Im}\chi_2(\mathbf{q},\omega,T_2) \quad (A.5)$$

**References**

[1] K.S. Novoselov, A. K. Geim, S.V. Morozov, D. Jiang, Y. Zhang, S. V. Dubonos, I.V. Grigorieva, A.A. Firsov, Science 306 (2004) 666.

[2] A.H. Castro Neto, F. Guinea, N.M.R. Peres, K.S. Novoselov, A.K. Geim, Rev. Mod. Phys. 81 (2009) 109.

[3] K.S. Novoselov, A. K. Geim, S.V. Morozov, D.Jiang, M.I. Katsnelson, I.V. Grigorieva, S.V. Dubonos, A.A. Firsov, Nature 438 (2005) 197.

[4] A.K. Geim, K.S. Novoselov, Nature Mater. 6 (2007) 183.

[5] S. Das Sarma, S. Adam, E. H. Hwang, E. Rossi, Rev. Mod. Phys. 83 (2011) 407.

[6] M.C. Lemme, Solid State Phenomena 156–158 (2010) 499.

[7] M.C. Lemme, T. Echtermeyer, M. Baus, H. Kurz, IEEE Electron Device Lett. 28 (4) (2007) 282.

[8] H. Schmidt, T. Ludtke, P. Barthold, E. McCann, V.I. Fal'ko, R.J. Haug, Appl. Phys. Lett. 93 (2008) 172108.

[9] M.Y. Kharitonov, K.B. Efetov, Phys. Rev. B 78 (2008) 241401.

[10] E.H. Hwang, S. Das Sarma, Phys. Rev. B 80 (2009) 205405.

[11] R.E.V. Profumo, M. Polini, R. Asgari, R. Fazio, A.H. McDonald, Phys. Rev. B 82 (2010) 085443.

[12] T. Vazifehshenas, T. Amlaki, M. Farmanbar, F. Parhizgar, Phys. Lett. A 374 (2010) 4899.

[13] A.G. Rojo, J. Phys.: Condens. Matter 11 (1999) R31.

[14] K. Flensberg, B.Y. Hu, Phys. Rev. B 52 (1995) 14796.

[15] N.P.R. Hill, J.T. Nichols, E.H. Linfeld, K.M. Brown, M. Pepper, D.A. Ritchie, G.A.C. Jones, B.Y. Hu, K. Flensberg, Phys. Rev. Lett. (1997) 78 2204.

[16] P.M. Solomon, P.J. Price, D.J. Frank, D.C.La Tulipe, Phys. Rev. Lett. 63 (1989) 2508.

[17] T.J. Gramila, J.P. Eisenstein, A.H. MacDonald, L.N. Pfeiffer, K.W. West, Phys. Rev. Lett. 66 (1991) 1216.

[18] U. Sivan, P.M. Solomon, H. Shtrikman, Phys. Rev. Lett. 68 (1992) 1196.

[19] H. Noh, S. Zelakiewics, X.G. Feng, T.J. Gramila, L.N. Pfeiffer, K.W. West, Phys. Rev. B 58 (1998) 12621.

[20] H.C. Tso, P. Vasilopoulos, F.M. Peeters, Phys. Rev. Lett. 68 (1991) 2516.

[21] A. Jauho, H. Smith, Phys. Rev. B 47 (1993) 4420.

[22] L. Zheng, A.H. MacDonald, Phys. Rev. B 48 (1993) 8203.

[23] A. Kamenev, Y. Oreg, Phys. Rev. B 52 (1994) 7516.

[24] L. Zheng, A.H. MacDonald, Phys. Rev. B 49 (1994) 5522.

[25] T. Vazifehshenas, A. Eskourchi, Physica E 36 (2007) 147.

[26] W-K Tse, B.Y-K Hu, S. Das Sarma, Phys. Rev. B. 76 (2007) 081401(R).

[27] M.I. Katsnelson, Phys. Rev. B. 84 (2011) 041407(R).

[28] S. Kim, I. Jo, J. Nah, Z. Yao, S.K. Banerjee, E. Tutuc, Phys. Rev. B 83 (2011) 161401(R).

[29] N.M.R. Peres, J.M.B. Lopes dos Santos, A.H. Castro Neto, Europphys. Lett. 95 (2011) 18001.

[30] D.K. Efimkin, Y.E. Lozovik, J. Exp.Theor. Phys. 113 (2011) 1050.

[31] E.H. Hwang, S. Das Sarma, Phys. Rev. B. 84 (2011) 255441.

[32] X.L. Lei, C.S. Ting, Phys. Rev. B 30 (1984) 4809.

[33] X.L. Lei X L, C.S. Ting, Phys. Rev. B 32 (1985) 1112.

[34] X.L. Lei, D.Y. Xing, M. Liu, C.S. Ting, Phys. Rev. B 36 (1987) 9134.

[35] C.S. Ting (Ed.), Physics of Hot Electron Transport in Semiconductor, World Scientific, Singapore, 1992.

[36] X.L. Lei, N.J.M. Horing, Int. J. Mod. Phys. B 6 (1992) 805.

[37] X.L. Lei, Phys. Status Solidi (b) 170 (1992) 519.

[38] X.L. Lei, N.J.M. Horing, H.L. Cui, J. Phys.: Condens. Matter 4 (1992) 9375.

[39] X.M. Weng, X.L. Lei, J. Phys.: Condens. Matter 6 (1994) 6287.

[40] H.S. Wu, M.W. Wu, Phys. Status Solidi (b) 192 (1995) 129.

[41] X.L. Lei, J.C. Cao, B. Dong, J. Appl. Phys. 80 (1996) 1504.

[42] H.S. Wu, M.W. Wu, X.X. Huang, Phys. Status Solidi (b) 198 (1996) 785.

[43] J.C. Cao, X. Lei, Commun. Theor. Phys .30 (1998) 381.

[44] X.L. Lei, Balance Equation Approach to Electron Transport in Semiconductors, World Scientific, Singapore, 2008.

[45] W.S. Bao, S.Y. Liu, X.L. Lei, C.M. Wang, J. Phys.: Condens. Matter 21 (2009) 305302.

[46] P.M. Solomon, B. Laikhtman, Superlattices Microstruct. 10 (1991) 89.

[47] C. Jacobini, P.J. Price, Solid State Electron. 31 (1988) 649.

[48] I.I. Boiko, Y.M. Sirenko, Phys. Stat. Sol. B 159 (1990) 805.

[49] B. Laikhtman, P.M. Solomon, Phys. Rev. B 41 (1990) 9921.

[51] B. Tanatar, J. App. Phys. 81 (1997) 6214.

[52] R.T. Senger, B. Tanatar, Solid State Commun. 121(2002) 61.

[53] S. Kasap, P. Capper (Eds.), Springer Handbook of Electronic and Photonic Materials, Springer, Berlin, 2006.